\documentclass[%
 reprint,
 amsmath,amssymb,
 aps, nofootinbib
]{revtex4-1}
\usepackage{graphicx}
\usepackage{dcolumn}
\usepackage{bm}

\usepackage{amsmath,amssymb}
\usepackage{slashed}
\usepackage{graphicx}
\usepackage{dsfont}
\usepackage{bm}
\usepackage[top=1.0in,bottom=1.0in,left=1.0in,right=1.0in]{geometry}
\usepackage[colorlinks,linkcolor=blue,citecolor=blue]{hyperref}
\usepackage{CJKutf8}
\usepackage{tikz-feynhand}

\begin{document}
\setlength{\oddsidemargin}{0.1cm}
\setlength{\topmargin}{-0.1cm}
\setlength{\textheight}{21cm}
\setlength{\textwidth}{17cm}
\newcommand{\be}{\begin{equation}}
\newcommand{\ee}{\end{equation}}
\newcommand{\bea}{\begin{eqnarray}}
\newcommand{\eea}{\end{eqnarray}}
\newcommand{\ba}{\begin{eqnarray}}
\newcommand{\ea}{\end{eqnarray}}

\newcommand{\fslash}{\hspace{-1.4ex}/\hspace{0.6ex} }
\newcommand{\Dslash}{D\hspace{-1.6ex}/\hspace{0.6ex} }
\newcommand{\Wslash}{W\hspace{-1.6ex}/\hspace{0.6ex} }
\newcommand{\pslash}{p\hspace{-1.ex}/\hspace{0.6ex} }
\newcommand{\kslash}{k\hspace{-1.ex}/\hspace{0.6ex} }
\newcommand{\underkslash}{{\underline k}\hspace{-1.ex}/\hspace{0.6ex} }
\newcommand{\epslash}{{\epsilon\hspace{-1.ex}/\hspace{0.6ex}}}
\newcommand{\partslash}{\partial\hspace{-1.6ex}/\hspace{0.6ex} }

\newcommand{\nn}{\nonumber}
\newcommand{\Tr}{\mbox{Tr}\;}
\newcommand{\tr}{\mbox{tr}\;}
\newcommand{\ket}[1]{\left|#1\right\rangle}
\newcommand{\bra}[1]{\left\langle#1\right|}
\newcommand{\rhoraket}[3]{\langle#1|#2|#3\rangle}
\newcommand{\brkt}[2]{\langle#1|#2\rangle}
\newcommand{\pdif}[2]{\frac{\partial #1}{\partial #2}}
\newcommand{\pndif}[3]{\frac{\partial^#1 #2}{\partial #3^#1}}
\newcommand{\pbm}[1]{\protect{\bm{#1}}}
\newcommand{\avg}[1]{\left\langle #1\right\rangle}
\newcommand{\vnabla}{\mathbf{\nabla}}
\newcommand{\notes}[1]{\fbox{\parbox{\columnwidth}{#1}}}
\newcommand{\pair}{\raisebox{-7pt}{\includegraphics[height=20pt]{pair0.pdf}}}
\newcommand{\paircrs}{\raisebox{-7pt}{\includegraphics[height=20pt]{pair0cross.pdf}}}
\newcommand{\paircc}{\raisebox{-7pt}{\includegraphics[height=20pt]{pair0cc.pdf}}}
\newcommand{\paircrscc}{\raisebox{-7pt}{\includegraphics[height=20pt]{pair0crosscc.pdf}}}
\newcommand{\pairloop}{\raisebox{-7pt}{\includegraphics[height=20pt]{pairloop.pdf}}}
\newcommand{\pairloopf}{\raisebox{-7pt}{\includegraphics[height=20pt]{pairloop4.pdf}}}
\newcommand{\pairlooph}{\raisebox{-7pt}{\includegraphics[height=20pt]{pair2looph.pdf}}}


\title{Nucleon Electric Dipole Form Factor in the QCD Instanton Vacuum}

\author{Wei-Yang Liu}
\email{wei-yang.liu@stonybrook.edu}
\affiliation{Center for Nuclear Theory, Department of Physics and Astronomy, Stony Brook University, Stony Brook, New York 11794-3800, USA}

\author{Ismail Zahed }
\email{ismail.zahed@stonybrook.edu}
\affiliation{Center for Nuclear Theory, Department of Physics and Astronomy, Stony Brook University, Stony Brook, New York 11794-3800, USA}

\date{\today}
\begin{abstract}
In the QCD vacuum, the nucleon form factors receive contributions from the underlying ensemble of topological pseudoparticles, which are sensitive to a finite vacuum angle $\theta$. We use this observation to derive a novel relationship between the Pauli and electric dipole form factors, for light quark flavors. This relationship allows for an explicit derivation of the proton and neutron electric dipole moments induced by a small CP violating $\theta$ angle, in terms of the vacuum topological susceptibity times pertinent magnetic moments. The results compare well with some recent lattice estimates.
\end{abstract}

\maketitle

\section{Introduction}
\label{SECI}
The resolution of the CP problem in baryogenesis is central to our understanding of baryon asymmetry in the universe~\cite{Sakharov:1967dj}. CP violation in the weak sector of the standard model, proves to be orders of magnitude away from its resolution, while CP violation in the strong sector, puts its resolution within range~\cite{Kharzeev:2019rsy}.

The QCD vacuum is rich with topologically active pseudoparticles
which are CP conjugated pairs (instantons and anti-instantons). They are natural sources of local CP violation effects. CP is strongly violated in QCD at finite theta angle. A reliable description of ensembles of these pseudoparticles in the semi-classical approximation, is provided by the instanton liquid model (ILM)~\cite{Diakonov:1995qy,Schafer:1996wv,Nowak:1996aj,Liu:2025ldh} (and references therein).
The model has proven to be very useful in capturing many aspects of most  hadronic correlations both in vacuum, in hadronic states and in matter. Here, we will focus 
on understanding the role of these pseudoparticles in the composition of the hadronic electric dipole moment.

For many decades, the nucleon electric dipole moment has been used as a measure  of the strong CP violation caused by a finite theta angle in QCD. The current empirical estimate puts its upper bound at about $10^{-26}\, e$-cm~\cite{Abel:2020pzs}. This is an ideal task for ab-initio lattice simulations, yet the smallness of the 
observable makes the task daunting in light of the signal-to-noise ratio~\cite{Syritsyn:2019vvt,Alexandrou:2020mds}. Notwithstanding this, the latest lattice estimate puts it at about
$10^{-16}\,\theta\,e$-cm for a finite theta angle~\cite{Liang:2023jfj}, which would limit $\theta$ to about $10^{-10}$ by the empirical bound.

The main purpose of this work is to evaluate the C-odd nucleon electric dipole form factor, and
to provide an estimation of the proton and neutron electric dipole moments in the ILM.  
We start by briefly reviewing the salient features of the pseudoparticle fluctuations in the ILM, with a comparison to some lattice simulations. 
With the definition of the four form factors associated to the electromagnetic current in the nucleon, the C-odd electric dipole form factor for each flavor contribution is evaluated in the ILM. This form factor is found to be related to the Pauli form factor. In particular, the finite proton and neutron electric dipole moments in the ILM are shown to be comparable to some recent lattice simulations. Further details can be found in Appendix~\ref{EMFF}

\begin{figure}
    \centering
    \includegraphics[width=.7\linewidth]{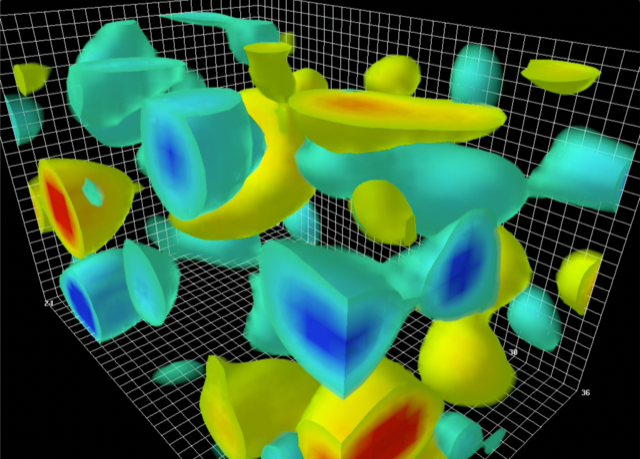}
    \caption{Visualization of instanton (yellow) and anti-instanton (blue) configurations in the deep-cooled Yang-Mills vacuum with~\cite{Moran:2008xq}.}
    \label{fig:inst_vac}
\end{figure}

\section{QCD instanton vacuum}
\label{SECII}

Recently, detailed gradient flow (cooling) techniques have revealed a striking semi-classical landscape made of instantons and anti-instantons, the vacuum tunneling configurations with unit topological charges~\cite{Leinweber:1999cw,Michael:1994uu,Michael:1995br,Biddle:2018bst,Athenodorou:2018jwu,Ringwald:1999ze}. These pseudoparticles in the QCD vacuum break conformal and chiral symmetry,
a mechanism governing most hadronic structure and dynamics. The key features of this landscape, as shown in Fig.~\ref{fig:inst_vac}, are characterized by the instanton plus anti-instanton density $n_{I+A}\equiv1/R^4\approx1/{\rm fm}^{4}$ and their average size $\rho\approx R/3$ respectively~\cite{Shuryak:1981ff}.
The hadronic scale $R=1\,{\rm fm}$
emerges as  the mean quantum tunneling rate of the pseudoparticles. Deep in the cooling time, the tunnelings are sparse, and well described by the instanton liquid model (ILM). For the details about the ILM and pertinent references, see Appendix~\ref{App:Inst_Vac}. 

\begin{figure}
    \centering
    \includegraphics[width=\linewidth]{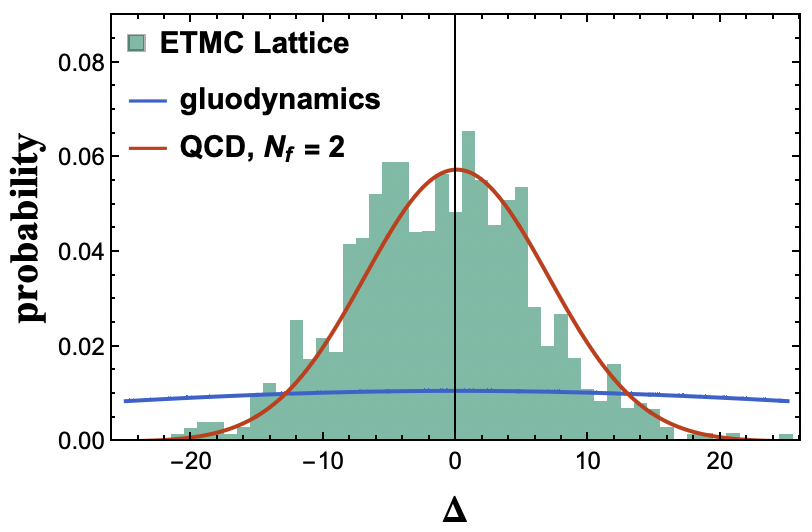}
    \caption{The ILM results of the topological charge $\Delta$ in unquenched (red-solid) and
    quenched (blue-solid) vacuum, compared to the lattice results from the ETMC collaboration~\cite{Alexandrou:2020mds} using twisted mass clover-improved fermions $N_f = 2+ 1+ 1$, in a 4-volume $64^3 \times 128~a^4$ with lattice spacing 
 $a=0.0801(4)$ fm, and physical pion mass $m_\pi=139$ MeV. For the ILM parameters see below.}
    \label{fig:topo-3}
 \end{figure}

To capture the salient features of the topological structures
in the vacuum with the number of pseudoparticles $N_\pm$, we use  the scalar $F^2_{\mu\nu}$ and pseudoscalar $F_{\mu\nu}\tilde{F}_{\mu\nu}$ gluonic densities in the vacuum. The vacuum expectation values of the gluonic densities at finite angle $\theta$, follow from low-energy theorems~\cite{Novikov:1981xi}
which are satisfied in the ILM.  In particular, the scalar and pseudoscalar densities are fixed by  the pseudoparticle density and the vacuum angle 
\bea
\label{SCALE}
\frac {\langle F^2\rangle_\theta}{32\pi^2}
&\approx& n_{I+A}\,{{\rm cos}\,\theta}\nonumber\\
\frac {\langle F\tilde F\rangle_\theta}{32\pi^2}
&\approx& in_{I+A}\,{{\rm sin}\,\theta}
\eea
to leading order in the density of pseuparticles. It receives corrections at higher orders  which are subleading in the dilute approximation. At this resolution, the scalar and pseudoscalar densities count the number of pseudoparticles. (See Appendix \ref{App:Inst_Vac} and \ref{App:grand})

In the absence of interactions, the fluctuations of the number sum of the pseudoparticles is given by a Poisson distribution. However, the interactions enhance the fluctuations with a distribution of their number stronger than Poisson. This amounts to a variance  per pseudoparticle equal to $4/b$, where $b=\frac{11}3N_c-\frac23 N_f$ is the one-loop coefficient stemming from the beta function, as originally suggested in~\cite{Schafer:1996wv,Diakonov:1995qy}. 

The ensemble of the gauge configuration of the vacuum creates a distribution of topological charge, as presented in Fig.~\ref{fig:topo-3}. In QCD vacuum, the topological charge is defined by
\begin{equation}
    \Delta=\frac1{32\pi^2}\int d^4x F^a_{\mu\nu}\tilde{F}^a_{\mu\nu}
\end{equation}
or the number difference of pseudoparticles $\Delta=N_+-N_-$. The fluctuations of the topological charge $\Delta$ 
is captured by the 
topological susceptibity $\chi_t$~\cite{Zahed:2021fxk,Diakonov:1995qy}. 
In the quenched ILM, the  fluctuations in $\Delta$ 
are Gaussian with broad width~\cite{Diakonov:1995qy,DelDebbio:2004mc}. However, in the present of $N_f$ light quark flavors of equal current mass $m$, these fluctuations are substantially screened, with a variance per mean pseudoparticle number $\bar N$ given by~\cite{Liu:2023yuj,Liu:2023fpj,Liu:2024rdm,Faccioli:2001ug,Pobylitsa1989TheQP}
\bea
\label{SUS}
 \chi_t=\frac{\langle \Delta^2\rangle}{V}\sim 
 n_{I+A}\bigg(1+N_f\frac {m^*}{m}\bigg)^{-1}
 \eea 
$V$ is the 4-volume of the vacuum 
and \textcolor{black}{$m^*$ is the determinantal mass (the effective mass defined by the gauge field configuration average of the fermionic determinant in the UV smeared lattice ensemble). The details of the determinantal mass  $m^*$ are given in Appendix~\ref{App:mass_det}. The value of $m^*$ for each quark is fixed by the chiral condensate, with an early rough estimate given in~\cite{Vainshtein:1981wh}
\begin{equation}
\label{mdet_gap}
    m^*\simeq m-\frac{2\pi^2\rho^2}{N_c}\langle \bar qq \rangle
\end{equation}
in leading order in the pseudoparticle density, with the quark condensate (for $q=u$ or $d$) given by
\begin{equation}
\begin{aligned}
\label{qq_leading}
    \langle \bar qq\rangle
    \simeq&-\frac{n_{I+A}}{m^*}+\mathcal{O}(n_{I+A}^2)
\end{aligned}
\end{equation}
The more detailed analysis in Appendix~\ref{App:mass_det} yields the value 
of the light condensate
$$\langle\bar \psi \psi\rangle=\langle\bar uu+\bar dd\rangle$$ quoted 
in Table~\ref{tab:parameters_ILM_2}, using the ILM parameters listed in Table~\ref{tab:parameters_ILM}.
Note that the rough estimate \eqref{qq_leading} gives instead $m^*=173$ MeV and $\langle\bar \psi \psi\rangle=(206~\mathrm{MeV})^3$, roughly of the same order as the ones presented in Table~\ref{tab:parameters_ILM_2}. The current mass is chosen such that the renormalization group (RG) invariant combination $m\langle\bar\psi\psi\rangle$, is within the consistent range compared to the lattice in \cite{FlavourLatticeAveragingGroupFLAG:2021npn}.}
 

\textcolor{black}{The result for $m^*$ estimated by \eqref{mdet_gap} is presented in Table~\ref{tab:parameters_ILM}, in consistent with the model estimation in \cite{Liu:2024rdm,Liu:2025ldh} and numerical simulation in \cite{Faccioli:2001ug}. The detail evaluation can also be found in Appendix~\ref{App:mass_det}. Using \eqref{SUS}, we now land the ILM prediction for topological susceptibility $\chi_t=(73.19~\mathrm{MeV})^4$.}

\textcolor{black}{We show the sensitivity of the topological distribution to the presence of quarks in  Fig.~\ref{fig:topo-3} and compare the lattice results from the ETMC collaboration~\cite{Alexandrou:2020mds} with three light flavors $N_f=2$ and current quark mass $m=6.0$ MeV, to the ILM results in the quenched (gluodynamics, solid-blue curve) and the 
unquenched (solid-red) ensemble using a Gaussian distribution with variance given by (\ref{SUS}). The parameters we used here is listed in Table~\ref{tab:parameters_ILM}.}

\begin{table}
    \centering
    \begin{tabular}{c|c|c|c}
     & $m$ &  $|\langle\bar{\psi}\psi\rangle|$ & $m|\langle\bar{\psi}\psi\rangle|$   \\[5pt] 
    \hline
    ILM & $6$ MeV & ($223.5$ MeV)$^3$ & $6.69\times10^{-5}$ \\
     \hline
    FLAG & $3.381(40)$ MeV & ($272(5)$ MeV)$^3$ & $6.80\times10^{-5}$ \\
   \hline
\end{tabular}
    \caption{Chiral condensate in the ILM using the parameters listed in Table~\ref{tab:parameters_ILM} at the resolution
    $\mu=1/\rho\approx600$ MeV. We compare the ILM results to the Flavour Lattice Averaging Group (FLAG) \cite{FlavourLatticeAveragingGroupFLAG:2021npn}
    at the resolution $\mu=$2 GeV. The last RG invariant combination is in GeV$^4$.}
    \label{tab:parameters_ILM_2}
\end{table}

\begin{table}
    \centering
    \begin{tabular}{c|c|c|c}
     $\rho$ & $n_{I+A}$ & $m$ & $m^*$   \\[5pt] 
    \hline
   $0.313$ fm & 1 fm$^{-4}$ & $6$ MeV & $103.6$ MeV \\
   \hline
\end{tabular}
    \caption{ILM parameters used in this work.}
    \label{tab:parameters_ILM}
\end{table}

 
\section{Nucleon EM form factors}
\label{SECIII}
At finite vacuum angle $\theta$, the QCD ground vacuum is CP violating.
In the ILM, this is seen by noting that the added $\theta$ term acts as a chemical potential for the topological charge $\Delta$, therefore enhancing instantons and depleting anti-instantons. This affects most observables, and in particular the  electromagnetic (EM) form factors of hadrons.

\begin{figure}
    \centering
    \includegraphics[scale=0.8]{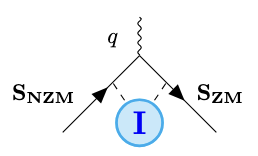}
    \caption{Leading pseudoparticle (single instanton) contribution to the quark EM current in the ILM. The dashed lines refer to the exchange induced by a classical instanton and a propagating quark. The details can be found in Appendix \ref{app:q_prop} and \ref{App:EM_inst}}
    \label{fig:VERTEX}
\end{figure}

In general, the nucleon-photon coupling can be described by Dirac $F_1$, Pauli $F_2$, the electric dipole moment form factor $F_3$, and the axial tensor form factor $F_A$
\begin{widetext}
\begin{equation}
\begin{aligned}
\label{EM_form}
    &\langle N(P')|J^{\mu}_{\rm EM}=\sum_f Q_f\bar{\psi}_f\gamma^\mu\psi_f |N(P)\rangle\\
    &=\bar{u}_{s'}(P')\left[\gamma^\mu F_1(Q^2)+\frac{i\sigma^{\mu\nu}q_\nu}{2M_N}\left(F_2(Q^2)-i\gamma^5F_3(Q^2)\right)+\frac{1}{M_N^2}\left(\slashed{q}q^\mu-q^2\gamma^\mu\right)\gamma^5F_A(Q^2) \right]u_{s}(P)
    \end{aligned}
\end{equation}
\end{widetext}
with 
$F_3$ and $F_A$ vanishing for $\theta=0$. 
The ensuing magnetic and electric dipole moments are,
\begin{align}
\mu_N=&e(F_{1}(0)+F_{2}(0))/2M_N \\
d_N=&eF_3(0)/2M_N
\end{align}
In Fig.~\ref{fig:VERTEX} we show how a single pseudoparticle (instanton or anti-instanton)  affects the electromagnetic vertex in the ILM. A light quark propagating in the instanton liquid background undergoes scattering into a zero mode (ZM) or a non-zero-mode (NZM) states, the eigenmodes of a Dirac operator. The ZMs are left handed in an instanton and right-handed in an anti-instanton~\cite{Liu:2024rdm,Liu:2025ldh,Zahed:2022wae}. This chirality selection rule implies that the EM vector-like vertex can only couple light quarks undergoing a ZM-NZM transition or vice-versa. This interaction induces a non-perturbative helicity-flip magnetic type coupling~\cite{Kochelev:2003cp}.

The explicit forms of the ZM and NZM  propagators are well-known~\cite{Schafer:1996wv} (see also in \eqref{ZMODE} and \eqref{NZM} in Appendix \ref{app:q_prop}). With this in mind and using the  
on-shell reduction scheme developed in~\cite{Liu:2021evw},
the instanton plus anti-instanton contributions presented in Fig.~\ref{fig:VERTEX}
modify the vector current operator for a single flavor
\begin{widetext}
\begin{equation}
\label{VPM}
  \frac1V\int d^4x e^{-iq\cdot x}\bar\psi_f(x)\gamma^\mu\psi_f(x) \rightarrow 
  \int\frac{d^4k}{(2\pi)^4}\frac{d^4k'}{(2\pi)^4}
  \left\langle\bar{\psi}_f(k')\left[\frac{N_+}{V}V^\mu_+(k',k)+\frac{N_-}{V}V^\mu_-(k',k)\right]\psi_f(k)\right\rangle
\end{equation}
with the vertex for the Pauli contribution
\bea
\label{V_inst}
    V^\mu_\pm(k',k)
    &\simeq&~ (2\pi)^4\delta(k-k'+q)\frac{8\pi^2\rho^4}{N_c}
\frac{1\mp\gamma^5}{2}\frac{i\sigma^{\mu\nu}q_\nu}{2m^*}\nonumber\\
&&\times\int_0^1dt\left[tK_0(u\sqrt{1-t})-\frac{1}{8}\frac{\sqrt{1-t}}{u}\frac{\partial}{\partial u}\left(uK_1(u\sqrt{1-t})\right)\right]\bigg|_{u=\rho Q}
\eea
\end{widetext}
where $K_n(x)$ is a  modified Bessel function of the second kind. 
More details of this derivation can be found in Appendix~\ref{App:EM_inst}.

At zero vacuum angle, 
the Pauli form factor $F_2$ 
in~\eqref{EM_form} receives a large contribution from pseudoparticles in the QCD instanton vacuum~\cite{Kochelev:2003cp}. This observation carries to the  CP-odd contribution $F_3$  at finite vacuum angle. For that, we need to average the EM form factor over the numbers of pseudoparticles  $N_\pm$ in the ILM. This procedure is by now standard~\cite{Diakonov:1995qy,Schafer:1996wv,Zahed:2021fxk},
with a more detailed review given in~\cite{Liu:2024rdm,Liu:2025ldh}. Some key steps can also be found in Appendix \ref{App:grand} and \ref{App:EM_inst}. 


\textcolor{black}{For simplicity, we model the nucleon by a simple description of a quark–scalar diquark compound, e.g., the proton with a spin up is $(uud)_\uparrow\approx u_\uparrow[ud]_0$. This schematic description captures certain part of the correlations in the nucleon, but not the full set~\cite{Schafer:1996wv}. Hence, the quark contribution to the proton spin structure is mostly carried through the unpaired $u$-quark.}

With this in mind, the Pauli form factor of a quark of flavor $f$
in a nucleon ($u$ for proton and $d$ for neutron), can be identified from (\ref{VPM}-\ref{V_inst}) by comparison and after inserting (\ref{SUS}), the result gives
\begin{widetext}
\bea
\label{EDM5}
\langle q_f(k')|J^{\mu}_{\rm EM} |q_f(k)\rangle_{\rm Pauli}
\simeq
Q_fe\bar{u}_s(k')F_I(\rho Q)\left(1 -
\frac{\langle\Delta^2\rangle}{\bar N}\theta\,
i\gamma^5\right)\frac{i\sigma_{\mu\nu}q_\nu}{2m^*} u_s(k)
\eea
\end{widetext}
The pseudoparticle  induced form factor $F_I(\rho Q)$ is related to \eqref{V_inst}. As its detailed form will not be important for the rest of our arguments, it is shown in \eqref{EDM6} in Appendix~\ref{App:EM_inst}.


\begin{figure}
    \centering
    \includegraphics[width=\linewidth]{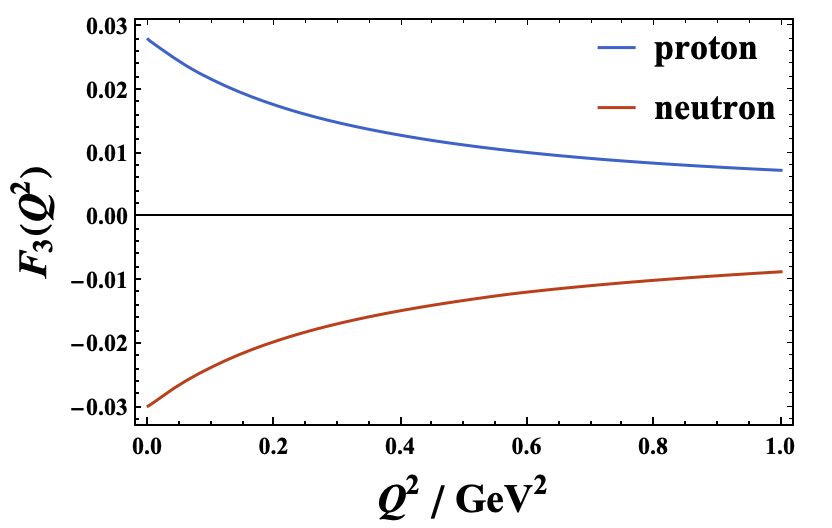}
    \caption{The CP-odd electric dipole form factor $F_3(Q^2)$ from the ILM for proton (blue) and neutron (red) is obtained using \eqref{F32F} with ILM parameters given in Table \ref{tab:parameters_ILM} and combining the parameters in \eqref{F32F} with the measurement \cite{Perdrisat:2006hj} for $F_2(Q^2)$ of proton and neutron fitted by dispersion analysis \cite{Belushkin:2006qa}.}
    \label{fig:F2}
\end{figure}

\begin{figure}
    \centering
    \includegraphics[width=\linewidth]{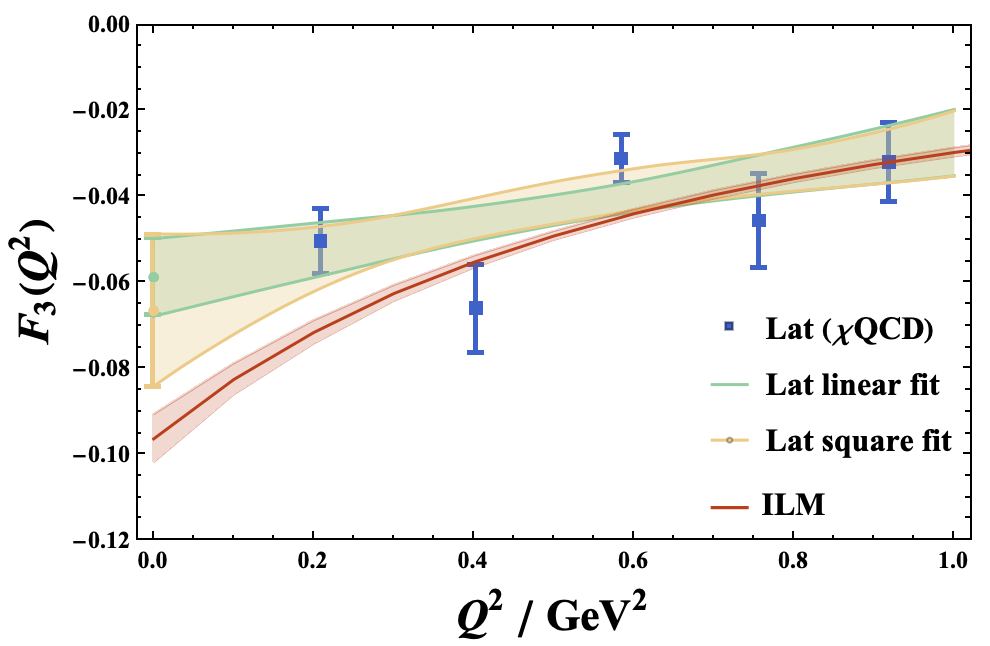}
    \caption{The neutron electric dipole form factor predicted using the ILM relation \eqref{F32F} combined with $F_2$ lattice results obtained by CSSM and QCDSF/UKQCD Collaborations \cite{CSSM:2014knt} in a 4-volume $32^3 \times64$ ensemble with pion mass $m_\pi=310$ MeV (red band) and lattice spacing $a=0.074(2)$ fm. 
    The $Q^2$ dependence is compared to the lattice results from $\chi$QCD collaboration~\cite{Liang:2023jfj} with $m_\pi=339$ MeV (blue data). The green band is a lattice 
    linear fit and the yellow band is a lattice 
    square fit with additional $Q^4$ terms.}
    \label{fig:enter-label}
\end{figure}

A comparison of (\ref{EDM5}) to (\ref{EM_form}) yields the pseudoparticle contribution to the $F_{2,3}$ $f$-quark form factors in leading order in the pseudoparticle density.
\bea
F^f_2(Q^2)&\rightarrow &Q_feF_I(\rho Q)\nonumber\\
    F^f_3(Q^2)&\rightarrow &Q_fe F_I(\rho Q)\frac{\langle\Delta^2\rangle}{\bar N}\theta
\eea
  
This connects the $f$-quark Pauli form factor to its electric dipole moment form factor
\begin{equation}
    \begin{aligned}
    \label{F32F}
    \frac{F^f_3(Q^2)}{ \theta}=&\frac{\chi_t}{n_{I+A}}F^f_2(Q^2)
    \end{aligned}
\end{equation}
for a small vacuum angle $\theta$ where the topological susceptibility $\chi_t$ is defined in \eqref{SUS}.
(\ref{F32F}) provides for a
relationship between the CP odd and even EM form factors.

\begin{table*}
\begin{center}
\caption{proton and neutron EDM.}
\begin{tabular}{|l|l|l|c|}
    \hline
    & Neutron ($10^{-3}~e\theta$$\cdot\rm fm$) & Proton ($10^{-3}~e\theta$$\cdot\rm fm$) & Ratio $|d_n/d_p|$ \\
    \hline
    ILM  & $d_n=-3.81$ & $d_p=3.57$ & $1.067$  \\
    Faccioli et al. \cite{Faccioli:2004jz}  & $|d_n|=6\sim14$ & $-$ & $-$  \\
    ChPT \cite{Mereghetti:2010kp} & $|d_n|=2.10$ & $|d_p|=2.38$ & $0.882$ \\
    $\chi$QCD\cite{Liang:2023jfj} & $d_n=-1.48^{+0.14}_{-0.31}$ & $d_p=3.8^{+1.1}_{-0.8}$ & $0.39^{+0.12}_{-0.12}$  \\
    Bhattacharya et al.\cite{Bhattacharya:2021lol} & $d_n=-3^{+7}_{-20}$ & $d_p=24^{+10}_{-30}$ & $0.13^{+0.80}_{-0.30}$ \\
    Dragos et al. \cite{Dragos:2019oxn} & $d_n=-1.52\pm 0.71$ & $d_p=1.1 \pm 1.0$ & $1.4 \pm 1.4$ \\
    ETMC\cite{Alexandrou:2020mds} & $|d_n|=0.9\pm2.4$ & $-$ & $-$  \\
    \hline
\end{tabular}
\label{TAB_1}
\end{center}
\end{table*}

Finally, using again (\ref{F32F}) at zero momentum transfer, we arrive at
a direct relationship between the nucleon dipole moments and the nucleon magnetic
moments
\begin{widetext}
\bea
\label{DNDP}
\frac{d_p(0)}{\theta}&=&\frac{\chi_t}{n_{I+A}}\times\left(\mu_p-Q_p\frac{e}{2M_N}\right)\nonumber\\
\frac{d_n(0)}{\theta}&=&\frac{\chi_t}{n_{I+A}}\times\left(\mu_n-Q_n\frac{e}{2M_N}\right)
\eea
\end{widetext}
where $Q_p=1$ and $Q_n=0$ is the charge of protons and neutrons.
In this section, (\ref{EDM5}), (\ref{F32F}) and (\ref{DNDP}) are the new and main results of this work.

\section{Results}
\label{SECIV}
In Fig.~\ref{fig:F2}, we present the electric dipole form factor $F_3(Q^2)$ predicted by the ILM at low $Q^2$, by using \eqref{F32F} with  $F_2(Q^2)$ borrowed from the dispersion analysis in~\cite{Perdrisat:2006hj} of the proton and neutron form factor measurements in~\cite{Belushkin:2006qa}.

In Fig.~\ref{fig:enter-label} (red band), we use \eqref{F32F} to derive the CP odd $F_3$, from the CP even lattice $F_2$ in~\cite{CSSM:2014knt} in a 4-volume $32^3 \times64$, with a pion mass $m_\pi=310$ MeV and lattice spacing $a=0.074(2)$ fm. The results are compared to the recently reported 
lattice results from the  $\chi$QCD collaboration (blue squares). \textcolor{black}{We choose the ILM parameters $m$ and $m^*$ for Fig.~\ref{fig:enter-label} by extending the physical point $(m=6~\mathrm{MeV},m_\pi=139~\mathrm{MeV})$ to the lattice choice of unphysical mass $(m=28.3~\mathrm{MeV},m_\pi=310~\mathrm{MeV})$ MeV with Gell-Mann–Oakes–Renner relation, assuming other low energy parameters such as $n_{I+A}$, $\rho$, and pion decay constant are insensitive to quark mass. The new determinantal mass $m^*$ is $142.4$ MeV, determined by Eq.~\eqref{mdet_gap_1} and \eqref{qq_1}.} The green band is a lattice linear fit, and the yellow band is a lattice square fit with additional $Q^4$ terms. Our ILM prediction ($m_\pi=310$ MeV) shows consistency with the recent lattice result in $\chi$QCD \cite{Liang:2023jfj} ($m_\pi=339$ MeV).

Using the empirical  values $\mu_p=2.793\, e/2M_N\simeq0.2937$ e$\cdot$fm and $\mu_n=-1.913\ e/2M_N\simeq-0.2012$ e$\cdot$fm~\cite{ParticleDataGroup:2016lqr},
 the ILM predicted electric dipole moments are listed in Table~\ref{TAB_1},
for the ILM parameters $N_f=2$, $m=6.0~\mathrm{MeV}$ and  $m^*= 103.6~\mathrm{MeV}$. One should keep in mind that $d_{p,n}$ in \eqref{DNDP} can be directly determined by the values of topological susceptibility and instanton density $n_{I+A}$ measured by experiments or lattice computations as well. The results are compared to the recently reported lattice results and chiral perturbation calculation (ChPT).  Overall, our results appear to be in the range of the proton and
neutron EDM reported by some lattice collaborations. For completeness, we note an earlier  estimate of the neutron dipole moment $|d_n(0)|=(6-14)\times10^{-3}\,(e\theta\cdot\rm fm)$ using  a numerical ensemble of pseudoparticles to describe the ILM, with a moment approximation to extract the EDM~\cite{Faccioli:2004jz}.

\textcolor{black}{In the current analysis of the ILM, the parameters entering our estimation of the nucleon EDM are fixed in Table \ref{tab:parameters_ILM_2} and \ref{tab:parameters_ILM}. Corrections to the present results are expected to arise from: 1/ omitted instanton-anti-instanton molecules; in the deep cooling regime where most of the UV configurations are smeared and the corresponding renormalization scale is a few hundred MeV, their contribution compared to the single-instanton one (current work) is about 10\% \cite{Liu:2025ldh}; 2/ deviations from the quark-diquark approximation used here. To address 1/ and 2/ thoroughly goes beyond the scope of this work.}

\section{Conclusions}
\label{SECV}
In QCD, the breaking of conformal symmetry puts stringent constraints on the bulk hadronic correlations in the form of low energy theorems~\cite{Novikov:1981xi}. These constraints are enforced in the QCD instanton vacuum in the form of stronger than Poisson fluctuations in the number of pseudoparticles, with a vacuum compressibility 
indicative of a quantum liquid. The fluctuations in the difference of the pseudoparticles are peaked around a neutral topological charge, with the variance fixed by the topological susceptibility. The latter 
is large in gluodynamics, but substantially screened in QCD with the presence of light quarks. These are essential features of the QCD vacuum as captured by the ILM.

In the ILM, the Pauli  form factor of  a constituent quark, receives a large contribution as noted initially in~\cite{Kochelev:2003cp}. In the presence of a  CP violation by a small vacuum angle $\theta$, we have shown that a CP-odd contribution develops in the Pauli form factor, driven mostly by the vacuum topological susceptibility. We have used this result to derive a 
"model-independent" relationship between the CP even $F_2$ and 
CP odd $F_3$ EM form factors, and to estimate the electric dipole moment of
both the  proton and neutron in terms of their respective magnetic moments. Our results are in the range of recently reported lattice results extrapolated at the physical pion mass~\cite{Liang:2023jfj}. The new relationship between $F_2$ and $F_3$ we derived, should prove useful for more accurate  lattice estimates of the nucleon dipole form factor.

\vskip 1cm
{\bf Acknowledgments\,\,}
We thank Fangcheng He for discussions. 
This work is supported by the Office of Science, U.S. Department of Energy under Contract  No. DE-FG-88ER40388.
This research is also supported in part within the framework of the Quark-Gluon Tomography (QGT) Topical Collaboration, under contract no. DE-SC0023646.

\appendix

\section{Nucleon EM form factors}
\label{EMFF}
Here we list the standard relationships between the commonly used Sachs form factors $G_{C,M}$ and $G_{D,A}$, and  the Dirac $F_1$, Pauli $F_2$, electric dipole moment $F_3$, and axial tensor $F_A$ form factors used in \eqref{EM_form}. 
\bea
\label{FFEM}
G_C(Q^2)&=&F_{1}(Q^2)+\frac{Q^2}{4M^2_N}F_2(Q^2) \\
G_M(Q^2)&=&F_{1}(Q^2)+F_{2}(Q^2) \\
G_D(Q^2)&=&F_{3}(Q^2)\\
G_A(Q^2)&=& F_A(Q^2)
\eea
The magnetic and electric dipole moment are defined respectively, as
\begin{align}
    \mu_N=G_M(0)e/2M_N && d_N=F_3(0)e/2M_N
\end{align}

\section{Instanton vacuum}
\label{App:Inst_Vac}
Quantum chromodynamics (QCD), the fundamental theory describing the strong interactions of quarks and gluons, possesses a rich and intricate vacuum structure characterized by topologically active gauge configurations, as illustrated in Fig.~\ref{fig:inst_vac}. The underlying nontrivial topological configurations play a crucial role in understanding the non-perturbative aspects in hadrons. 

The simplest topological gauge configuration carrying one unit of topological charge is a QCD instanton, the self-dual configuration corresponding to a zero energy tunneling event between two different vacuum states with Chern-Simon winding number changed by one unit. In singular gauge, the instanton gauge field is given by 

\begin{widetext}
\begin{equation}
\label{FSTX}
A_{\mu}(x;z,\rho,U)=-\frac1{2i}U\left(\tau_\mu^-\tau_\nu^+-\tau_\nu^+\tau_\mu^-\right)U^\dagger\partial_\nu\ln\left(1+\frac{\rho^2}{(x-z)^2}\right)
\end{equation}
\end{widetext}
with its center located at $z$, its size characterized by $\rho$, and its orientation in color space determined by $U$ ($SU(N_c)$ group element), reflecting the underlying classical symmetries of translational, scale, and color in the vacuum. $\tau^\pm_\mu=(\vec{\tau},\mp i)$ is defined in Euclidean signature with the $SU(N_c)$ generalized Pauli matrices $\vec{\tau}$ defined as an $N_c\times N_c$ matrix with $2\times2$ Pauli matrices embedded in the upper left corner. For the anti-instanton field, we interchange the Pauli matrices  $\tau^+_{\mu}\leftrightarrow\tau^-_{\mu}$. 

Phenomenologically, we can define \( N_+ \) as the number of instantons in the vacuum and \( N_- \) as the number of anti-instantons (instantons with opposite topological charges). The topological gauge  configurations in the vacuum can be sampled by those instantons as pseudoparticles in a 4-dimensional Euclidean box with residual moduli interactions, providing the most compelling representation of the underlying gauge configurations at low resolution. The vacuum tunneling probability per unit volume is well described by a semi-empirical formula given by
\begin{equation}
\label{dn_dist}
n(\rho) \sim  {1 \over \rho^5}\big(\rho \Lambda_{QCD} \big)^{b} \, e^{-C\rho^2/R^2}
\end{equation}
Here $b=11N_c/3-2N_f/3$ is the one loop beta function obtained by integrating over the perturbative gauge fields around the non-trivial instanton vacuum,
and $C$ is a number of order 1 fixed by the color interactions among pseudoparticles in the vacuum~\cite{Diakonov:1995ea,Shuryak:1999fe}. 

The number of these pseudoparticles $N_\pm$ are identified with the scalar and pseudoscalar gluonic operator in the vacuum at the leading order in density of pseudoparticles.
\bea
\label{NPM}
&&\frac {1}{32\pi^2}\int d^4x\,F_{\mu\nu}^2(x)= (N_++N_-)=N\nonumber\\
&&\frac {1}{32\pi^2}\int d^4x\,F_{\mu\nu}\tilde{F}_{\mu\nu}(x)= (N_+-N_-)=\Delta
\eea
 and their fluctuations can be calculated by the gluonic operator correlation in the vacuum.  The fluctuations in the difference $\Delta$ (topological charge) is captured by a Gaussian distribution with the variance fixed by the topological susceptibity $\chi_t$~\cite{Diakonov:1995qy}.
At zero vacuum angle, the mean value $\langle \Delta\rangle$ is expected to be zero and the topological susceptibility in gluodynamics (quenched vacuum) is of order $1$
However, in QCD it is substantially screened by the light quarks.


\section{Instanton ensemble at finite angle}
\label{App:grand}

In the QCD vacuum filled with topologically active configurations, the vacuum expectation values of most quarks and gluon operators are averaged through
\begin{equation}
\label{grand_canonical}
    \langle \mathcal{O}\rangle=\sum_{N_+,N_-}\mathcal {P}(N_+,N_-)\langle \mathcal{O}\rangle_{N_\pm}
    \equiv \overline{\langle\mathcal{O}\rangle}_{N_\pm} 
\end{equation}
The  averaging is carried out over the configurations with fixed $N_\pm$ (canonical ensemble average), followed by an ensemble averaging over the distribution \eqref{DISTX}.

\begin{equation}
\label{DISTX}
  \mathds{\mathcal P}(N_+,N_-)=\left[e^{\frac{bN}4 }\bigg(\frac {\bar{N}}{N}\bigg)^{\frac {bN}4 }\right]\left[\frac{1}{\sqrt{2\pi\chi_t}}\exp\left(-\frac{\Delta^2}{2\chi_t}\right)\right]
\end{equation}

These gauge configuration yield a grand canonical ensemble parameterized by vacuum angle $\theta$ describing how favorable the vacuum to have positive topological charge and vacuum chemical potential $\mu$ describing the tendency to create a pseudoparticle in the vacuum, which is set to be zero since pseudoparticles do not carry energy~\cite{Diakonov:1995qy,Schafer:1996wv,Zahed:2021fxk}. With the ensemble distribution in \eqref{DISTX}, the grand canonical partition function $\mathcal{Z}$ can be reconstructed from the moments obtained from \eqref{DISTX}, capturing the vacuum dependence on the chemical potential $\mu$ and the vacuum angle $\theta$.

\begin{equation}
\begin{aligned}
    \mathcal{Z}(\mu,\theta)\equiv&\sum_{N_\pm}Z_{N_\pm}e^{(\mu+i\theta)N_++(\mu-i\theta)N_-}\\
    =&\exp\left[\frac b4\langle N\rangle_\theta \exp\left(\frac 4 b \mu\right)\right]
\end{aligned}
\end{equation}
where $Z_{N_\pm}$ is the canonical partition function with fixed number of pseudoparticles. 

Given the partition function $\mathcal{Z}$ in this ensemble, the angle dependence of $\langle N\rangle_\theta$ and $\langle \Delta\rangle_\theta$ in finite theta vacuum is determined by 

\begin{equation}
\label{N_theta}
\frac{\partial\langle N \rangle_\theta}{\partial \theta}=i\frac4b\langle \Delta \rangle_\theta
\end{equation}

At small finite angle, the solutions to the co-dependent equations \eqref{N_theta} are given by \cite{Liu:2025ldh,hutter2001instantonsqcdtheoryapplication}

\begin{align}
    \langle N\rangle_\theta=&\bar{N}+\mathcal{O}(\theta^2)\\
    \langle \Delta\rangle_\theta=& i
    \langle\Delta^2\rangle\,\theta+\mathcal{O}(\theta^3)
\end{align}

\section{Determinantal mass $m^*$}
\label{App:mass_det}
In the ILM, the concept of a determinantal mass $m^*$ 
characterises the width of the zero-mode zone (ZMZ), a key feature of the
spontaneous breaking of chiral symmetry. The determinantal mass is given by the average of the fermionic determinant in the instanton ensemble with $N$ pseudoparticles,
\begin{equation}
    \left\langle\prod_f\mathrm{Det}(\slashed{D}+m)\right\rangle=(m^*)^{NN_f}
\end{equation}
a measure of the unquenching of the gauge configurations. \textcolor{black}{If we only consider the contributions from zero modes and the lowest order in the quark mass $m$ expansion in the chiral limit, the ensemble average of the fermionic determinant yields  a gap-like equation \cite{Liu:2025ldh,Schafer:1996wv},
\begin{equation}
\label{mdet_gap_1}
    m^*\simeq m-\frac{2\pi^2\rho^2}{N_c}\langle \bar qq \rangle
\end{equation}
with the single species quark condensate 
\begin{equation}
\label{qq_1}
    \langle \bar qq \rangle= -4N_c\int\frac{d^4k}{(2\pi)^4}\frac{M(k)}{k^2+M^2(k)}\mathcal{F}(\rho k)
\end{equation}
The dynamical constituent quark mass $M(k)$ is given by
\begin{equation}
   M(k)=m+(M(0)-m)\mathcal{F}(\rho k)
\end{equation}
with $M(0)=395$ MeV fitted to the  light meson mass spectrum \cite{Schafer:1996wv,Kock:2020frx,Liu:2023yuj,Liu:2023fpj,Liu:2025ldh}, and $\mathcal{F}(k)$ is the profile of the quark zero mode~\cite{Liu:2023yuj,Kock:2020frx} defined by
\begin{equation}
    \sqrt{\mathcal{F}(k)}=-z\frac{d}{dz}[I_0(z)K_0(z)-I_1(z)K_1(z)]\bigg|_{z=\frac{\rho k}{2}}
\end{equation}}

\textcolor{black}{With this in mind, \eqref{mdet_gap_1} gives $m^*=103.6$ MeV, which is consistent with the 
result $m^*\sim 103\,\rm  MeV$~\cite{Faccioli:2001ug}, following from numerical simulations of ensembles of interacting pseudoparticles.}

In the ILM, the vacuum transition amplitude is 
\begin{equation}
    \frac{n_{I+A}}{2}=\int d\rho n(\rho)(m^*\rho)^{N_f}
\end{equation}
where $n(\rho)$ defined in \eqref{dn_dist} denotes the quenched instanton size distribution. The effect of the  determinantal mass $m^*$ is to quench the tunneling effects. The numerical value for the ILM parameters combination
\begin{equation}
    \frac{n_{I+A}}{2}
    \left(\frac{4\pi^2\rho^2}{m^*}\right)^{N_f}\approx610.3~\mathrm{GeV}^{-2}
\end{equation}
generate the correct pion mass~\cite{Liu:2023yuj}.

\section{Quark propagator in an instanton background}
\label{app:q_prop}
\begin{figure}
    \centering
    \includegraphics[width=0.4\linewidth]{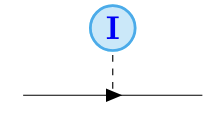}
    \caption{The illustration of the quark propagation with a single instanton background. The dashed line denotes the presence of the instanton (anti-instanton) as a background field for the propagating quark defined by $S_I$ in \eqref{eq:SIA_prop}}
    \label{fig:q_p}
\end{figure}
In the presence of an instanton (anti-instanton) configuration, the quark propagator is distorted by the instanton (anti-instanton) background field. \textcolor{black}{The quark propagator can be obtained by the spectral decomposition with the eigen states $\psi^I_\lambda$ defined by Dirac operator $\slashed{D}=\slashed{\partial}-i\slashed{A}_I$ with given instanton field configuration $A_I$.}

\begin{equation}
    i\slashed{D}\psi^I_\lambda=\lambda \psi^I_\lambda
\end{equation}

\textcolor{black}{The zero mode propagator corresponds to the contribution from the quark eigen state with zero eigenvalue $\lambda=0$ and non-zero mode propagator corresponds to the contribution from the quark eigen states with non-zero eigenvalues $\lambda\neq0$.} 
For small quark masses, the propagator can be expanded around the chiral limit ($m\rightarrow0$). Thus, the propagator with a single instanton field background $A_I$ defined in \eqref{FSTX}, can be expressed as
\begin{widetext}

\begin{equation}
\begin{aligned}
\label{eq:SIA_prop}
    S_I(x,y)=&\langle x|\frac{1}{i\slashed{\partial}+\slashed{A}_I+im}|y\rangle\\
    =&S_{\mathrm{ZM}}(x-z,y-z)+S_{\mathrm{NZM}}(x-z,y-z)+\mathcal{O}(m)
\end{aligned}
\end{equation}
where the quark propagation dependes on the center of the instanton $z$. This effective propagator can be illustrated by Fig.~\ref{fig:q_p} where the dashed line denotes the presence of the instanton (anti-instanton) as a background field for the propagating quark defined by $S_I$ in \eqref{eq:SIA_prop}

The zero mode propagator reads
\begin{equation}
\begin{aligned}
\label{ZMODE}
    S_{\mathrm{ZM}}(x,y)=&\frac{1}{im^*}
    \left[\frac{\rho^2}{8\pi^2}\frac{x_\alpha y_\beta}{x^2(x^2+\rho^2)y^2(y^2+\rho^2)}U\tau_\mu^-\tau_\nu^+U^\dagger\gamma_\alpha\gamma_\mu\gamma_\nu\gamma_\beta\right]\frac{1+\gamma^5}{2}\\
    &\times\frac1{(1+\rho^2/x^2)^{1/2}(1+\rho^2/y^2)^{1/2}}
\end{aligned}
\end{equation}
The singular $1/m$ in the single instanton zero modes has been shifted to finite $1/m^*$ in \eqref{ZMODE} due to the disordering in the multi-instanton background~\cite{Pobylitsa1989TheQP,Schafer:1996wv,Liu:2025ldh}.

The non-zero mode propagator in the chiral-split form reads \cite{PhysRevD.17.1583}
\begin{equation}
\begin{aligned}
\label{NZM}
    &S_{\mathrm{NZM}}(x,y)
    =S_{nz}(x,y)\frac{1+\gamma^5}{2}+\bar{S}_{nz}(x,y)\frac{1-\gamma^5}{2}
\end{aligned}
\end{equation}


where $S_{nz}$ and $\bar{S}_{nz}$ can be recast in the form~\cite{Schafer:1996wv,Zubkov:1997fn,Liu:2021evw}

\begin{equation}
\begin{aligned}
    S_{nz}(x,y)=&\left[\frac{-i(\slashed{x}-\slashed{y})}{2\pi^2(x-y)^4}\left(1+\rho^2\frac{x_\mu y_\nu}{x^2y^2}U\tau_\mu^-\tau_\nu^+U^\dagger\right)-\frac{\rho^2\gamma_\mu}{4\pi^2}\frac{x_\rho(x-y)_\nu y_\lambda}{(x^2+\rho^2)x^2(x-y)^2y^2}U\tau_\rho^-\tau^+_\mu\tau_\nu^-\tau_\lambda^+U^\dagger\right]\\
    &\times\frac{1}{\left(1+\rho^2/x^2\right)^{1/2}\left(1+\rho^2/y^2\right)^{1/2}}
\end{aligned}
\end{equation}
and
\begin{equation}
\begin{aligned}
    \bar{S}_{nz}(x,y)=&\left[\frac{-i(\slashed{x}-\slashed{y})}{2\pi^2(x-y)^4}\left(1+\rho^2\frac{x_\mu y_\nu}{x^2y^2}U\tau_\mu^-\tau_\nu^+U^\dagger\right)-\frac{\rho^2\gamma_\mu}{4\pi^2}\frac{x_\rho(x-y)_\nu y_\lambda}{(y^2+\rho^2)x^2(x-y)^2y^2}U\tau_\rho^-\tau^+_\nu\tau_\mu^-\tau_\lambda^+U^\dagger\right]\\
    &\times\frac{1}{\left(1+\rho^2/x^2\right)^{1/2}\left(1+\rho^2/y^2\right)^{1/2}}
\end{aligned}
\end{equation}

Note that the propagator for the anti-instanton can be obtained via the substitutions  $\tau^-_\mu\leftrightarrow\tau^+_\mu$, and $\gamma^5\leftrightarrow-\gamma^5$.
\\[10pt]
\section{Emergent EM vertex modified by single instanton}
\label{App:EM_inst}
To see how the pseudoparticles contribute to the electromagnetic form factors of hadrons, 
we consider the instanton (anti-instanton) insertion to the charge
current on a quark line.
In an instanton (anti-instanton) the incoming 
left-handed (right-handed) quark flips to a right-handed (left-handed) 
through a zero mode, then scatters off a
virtual photon before exiting through a non-zero mode, with the result for a single quark~\cite{Kochelev:2003cp}.  The effective interaction amplitude is defined as

\begin{equation}
\begin{aligned}
\label{VIIBAR} 
V^\mu_+(x,y)=\int d^4zdU\int d^4ye^{-iq\cdot y} &\bigg[S_{\mathrm{NZM}}(x-z,y-z)\gamma^\mu S_{\mathrm{ZM}}(y-z,x'-z)\\
&+ S_{\mathrm{ZM}}(x-z,y-z)\gamma^\mu S_{\mathrm{NZM}}(y-z,x'-z)\bigg]
\end{aligned}
\end{equation}
with the instanton center $z$ and the color orientation $U$ fixed.

Inserting the zero mode (\ref{ZMODE}) and nonzero mode \eqref{NZM} into (\ref{VIIBAR}) yields the instanton induced effective EM vertex. The effective vertex for the anti-instanton $V^\mu_-(x,y)$ follows by interchanging $\tau^-_\mu\leftrightarrow\tau^+_\mu$, $\sigma_\mu\leftrightarrow\bar{\sigma}_\mu$, and $\gamma^5\leftrightarrow-\gamma^5$.
With the on-shell reduction of the in-out quark lines, the effective quark operator modified by single instanton can be written as 

\begin{equation}
\begin{tikzpicture}[scale=0.5,baseline=(o)]
   \begin{feynhand}
   \path (0,0) -- (4,0);
    \node at (-0.5,1) {$\mathbf{S_{NZM}}$};
    \node at (4.1,1) {$\mathbf{S_{ZM}}$};
    \vertex (a) at (0,0) ;
    \vertex (c) at (4,0) ;
    \vertex (b) at (2,2);
    \vertex (d) at (2,3);
    \vertex (o) at (1,1);
    \vertex (g0) at (2,0.25);
    \vertex (g1) at (1.3,1.3);
    \vertex (g2) at (2.7,1.3);
    \vertex (g3) at (2.5,1.5);
    \vertex (g4) at (1.3,1.3);
    \vertex (g5) at (4.5,1);
    \vertex (g6) at (5,-1);
    \vertex (g7) at (7,-1);
    \vertex (g8) at (6,-1);
    \node at (1.3,2.4) {$q$};
    \propag [fer] (a) to (b);
    \propag [fer] (b) to (c);
    \propag [pho] (b) to (d);
    \propag [sca] (g0) to (g1);
    \propag [sca] (g0) to (g2);
    \filldraw[fill=cyan!1, color=cyan!20, draw=cyan, very thick] (2,0.25) circle (0.6);
    \node at (2, 0.25) {\textcolor{blue}{\bf \Large I}};
   \end{feynhand}
\end{tikzpicture}= \int\frac{d^4k}{(2\pi)^4}\frac{d^4k'}{(2\pi)^4}\bar{\psi}(k')\left[\frac{ N_+}{V}V^\mu_+(k',k)+\frac{N_-}{V}V^\mu_-(k',k)\right]\psi(k)
\end{equation}
where the momentum-dependent vertex function is obtained by on-shell reduction using the large time asymptotics, 
\begin{equation}
    \begin{aligned}
\label{vertice}
    V^\mu_+(k',k)=&\lim_{\substack{x_4\rightarrow+\infty\\
    y_4\rightarrow-\infty}}\left[\int d^3\vec{x} d^3\vec{y} e^{ik'\cdot x} i\gamma_4V^\mu_+(x,y)i\gamma_4 e^{-ik\cdot y}\right]_{\substack{ik_4=|\vec{k}|\\ ik'_4=|\vec{k}'|}}\\
    \end{aligned}
\end{equation}
More specifically, 
the on-shell reduction of the Euclidean and massless quark propagator in the instanton background, can be achieved through the large Euclidean time asymptotic 
\begin{equation}
\begin{aligned}
    &\lim_{\tau\rightarrow-\infty}i\int d^3\vec{y}S_{nz}(x,y)e^{i\vec{k}\cdot\vec{y}}\left[\gamma_4\frac{1-\gamma^5}2 e^{-|\vec{k}|\tau}\right]\\
    =&\frac{e^{-ik\cdot x}}{(1+\rho^2/x^2)^{1/2}}\Bigg[1-i\frac{\rho^2}{2x^2}  x_\mu k_\nu U\tau^-_\mu\tau^+_\nu U^\dagger\int_0^1dtte^{i(1-t)k\cdot x}\\
    &+i\frac{\rho^2}{4x^2(x^2+\rho^2)}\frac{ x_\mu\gamma_\nu\gamma_4}{|\vec{k}|} x_\rho k_\lambda U\tau^-_\mu\tau^+_\nu\tau^-_\rho\tau^+_\lambda U^\dagger\int_0^1dte^{i(1-t)k\cdot x}+\frac{\rho^2}{2x^2(x^2+\rho^2)}\frac{x_\mu\gamma_\nu\gamma_4}{|\vec{k}|} U\tau^-_\mu\tau^+_\nu U^\dagger\Bigg]\frac{1-\gamma^5}2
\end{aligned}
\end{equation}
and
\begin{equation}
\begin{aligned}
    &\lim_{\tau\rightarrow\infty}i\int d^3\vec{y}\left[\frac{1-\gamma^5}2\gamma_4e^{|\vec{k}|\tau}\right]\bar{S}_{nz}(y,x)e^{-i\vec{k}\cdot\vec{y}}\\
    =&\frac{e^{ik\cdot x}}{(1+\rho^2/x^2)^{1/2}}\frac{1-\gamma^5}2\Bigg[1+i\frac{\rho^2}{2x^2} k_\mu x_\nu U\tau^-_\mu\tau^+_\nu U^\dagger\int_0^1dtte^{-i(1-t)k\cdot x}\\
    &-i\frac{\rho^2}{4x^2(x^2+\rho^2)}\frac{\gamma_4\gamma_\nu}{|\vec{k}|} k_\lambda x_\rho x_\mu  U\tau^-_\lambda\tau^+_\rho\tau^-_\nu\tau^+_\mu U^\dagger\int_0^1dte^{-i(1-t)k\cdot x}+\frac{\rho^2}{2x^2(x^2+\rho^2)}\frac{\gamma_4\gamma_\mu x_\nu}{|\vec{k}|} U\tau^-_\mu\tau^+_\nu U^\dagger\Bigg]
\end{aligned}
\end{equation}
After a few algebraic simplification, the result for the ZM-NZM mixing contribution in the EM vertex reads
\bea
    V^\mu_\pm(k',k)
    &\simeq&~ (2\pi)^4\delta(k-k'+q)\frac{8\pi^2\rho^4}{N_c}
\frac{1\mp\gamma^5}{2}\frac{i\sigma^{\mu\nu}q_\nu}{2m^*}\nonumber\\
&&\times\int_0^1dt\left[tK_0(u\sqrt{1-t})-\frac{1}{8}\frac{\sqrt{1-t}}{u}\frac{\partial}{\partial u}\left(uK_1(u\sqrt{1-t})\right)\right]\bigg|_{u=\rho Q}
\eea
where $K_n(x)$ is the modified Bessel function of the second kind. Now the single quark matrix element 

\begin{equation}
\langle q_f(k')|J^{\mu}_{\rm EM} |q_f(k)\rangle
\simeq
Q_fe\bar{\psi}_f(k')F_I(\rho Q)\left[\frac{\langle N_+\rangle_\theta}{\bar N}\frac{1-\gamma^5}2+\frac{\langle N_-\rangle_\theta}{\bar N}\frac{1+\gamma^5}2\right]\frac{i\sigma_{\mu\nu}q_\nu}{2m^*}\psi_f(k)
\end{equation}
where the pseudo-particle induced form factor is defined as
\begin{equation}
\begin{aligned}
\label{EDM6}
    F_I(q)= \frac{8\pi^2\rho^4n_{I+A}}{N_c} \frac{1}{q^2}\left[\bigg(2-2qK_1(q)\bigg)-\left(\frac7{q^2}-\frac{7}{4}qK_1(q)-\frac72K_2(q)\right)\right]
\end{aligned}
\end{equation}

Taking the fluctuation into account, which gives $2\langle N_\pm\rangle_\theta=\bar{N}\pm i\langle\Delta^2\rangle\,\theta$ (See Appendix~\ref{App:grand}), the Pauli contribution to each single quark form factor is given by
\bea
\langle q_f(k')|J^{\mu}_{\rm EM} |q_f(k)\rangle
\simeq
Q_fe\bar{\psi}_f(k')\left[F_I(\rho Q)\left(1 -
\frac{\langle\Delta^2\rangle}{\bar N}\theta\,
i\gamma^5\right)\right]\frac{i\sigma_{\mu\nu}q_\nu}{2m^*} \psi_f(k)
\eea

\end{widetext}






\bibliography{EDM}

\end{document}